\documentclass{article}

\usepackage{arxiv}

\usepackage[utf8]{inputenc} % allow utf-8 input
\usepackage[T1]{fontenc}    % use 8-bit T1 fonts
\usepackage{hyperref}       % hyperlinks
\usepackage{url}            % simple URL typesetting
\usepackage{booktabs}       % professional-quality tables
\usepackage{amsfonts}       % blackboard math symbols
\usepackage{nicefrac}       % compact symbols for 1/2, etc.
\usepackage{microtype}      % microtypography
\usepackage{graphicx}
\usepackage{natbib}
\usepackage{doi}

\title{Assessing the Effects of Container Handling Strategies on Enhancing Freight Throughput}

\date{}

\author{ \href{https://orcid.org/0009-0005-1487-9970}{\includegraphics[scale=0.06]{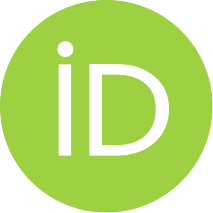}\hspace{1mm}Sarita Rattanakunuprakarn}\\
	Department of Industrial and System Engineering\\
	University of Tennessee, Knoxville\\
	Knoxville, TN 37996 \\
	\texttt{SRATTAN3@tennessee.edu} \\
	\And
    Mingzhou Jin \\
	Department of Industrial and System Engineering\\
	University of Tennessee, Knoxville\\
	Knoxville, TN 37996 \\
	\texttt{jin@utk.edu} \\
	\AND
	Mustafa Can Camur \\
	Department of Industrial and System Engineering\\
	University of Tennessee, Knoxville\\
	Knoxville, TN 37996 \\
	\texttt{mcamur@utk.edu} \\
	\And
	Xueping Li \\
	Department of Industrial and System Engineering\\
	University of Tennessee, Knoxville\\
	Knoxville, TN 37996 \\
	\texttt{xli27@utk.edu} \\
}

\renewcommand{\shorttitle}{Effects of Container Handling
Strategies}

\hypersetup{
pdftitle={A template for the arxiv style},
pdfsubject={q-bio.NC, q-bio.QM},
pdfauthor={David S.~Hippocampus, Elias D.~Striatum},
pdfkeywords={First keyword, Second keyword, More},
}

\begin{document}
\maketitle

\begin{abstract}
	As global supply chains and freight volumes grow, the U.S. faces escalating transportation demands. The heavy reliance on road transport, coupled with the underutilization of the railway system, results in congested highways, prolonged transportation times, higher costs, and increased carbon emissions. California's San Pedro Port Complex (SPPC), the nation's busiest, incurs a significant share of these challenges.
We utilize an agent-based simulation to replicate real-world scenarios, focusing on the intricacies of interactions in a modified intermodal inbound freight system for the SPPC. This involves relocating container classification to potential warehouses in California, Utah, Arizona, and Nevada, rather than exclusively at port areas. Our primary aim is to evaluate the proposed system's efficiency, considering cost and freight throughput, while also examining the effects of workforce shortages. Computational analysis suggests that strategically installing intermodal capabilities in select warehouses can reduce transportation costs, boost throughput, and foster resource balance in port complex areas.
\end{abstract}

% % keywords can be removed
% \keywords{First keyword \and Second keyword \and More}

\section{Introduction}
\label{sec:intro}

Intermodal transportation involves the integration of various transportation modes, such as trucks, trains, ships, and occasionally planes, using standardized containers.
This approach enhances flexibility, reduces transportation costs, and minimizes environmental impact compared to relying solely on a single mode of transportation. 
Despite shippers typically experiencing cost reductions ranging from 10\% to 40\% when transitioning freight traveling over 500 miles from highways to intermodal rail \citep{CSXIntermodal2023}, intermodal rail still faces underutilization. 
Inland intermodal transportation, consisting of rail and trucks, plays a crucial role in the efficiency of port operations, especially for major port complexes like the SPPC.
 
Comprising the Port of Los Angeles (POLA) and the Port of Long Beach (POLB), SPPC handled 29\% of all U.S. containerized international waterborne trade in 2021, representing 75\% of such trade in the West Coast region \citep{poLA2022factandfigure}. POLA has maintained its status as the U.S.'s leading container port for 23 years, from 2000 to 2022. However, this pivotal role comes with many urgent issues as explained below.

The port is grappling with a severe shortage of workforce, truck drivers, and chassis, leading to prolonged waiting times for vessels and trucks. On January 19, 2022, approximately 100 container vessels encountered substantial delays, spanning from 17.6 to 28 days, while awaiting clearance to dock at the port complex \citep{WOC2022vesselWaitTime} before gaining access to the SPPC. In some cases, vessels might be rerouted and unloaded at a different port, requiring additional transportation to bring the rerouted freight back to the original destination port \citep{McKenziePortOper}.

A primary concern revolves around the scarcity of truck drivers, as illustrated by 81,258 delays in 2021, with projections indicating a potential increase to approximately 160,000 by 2031 \citep{ATA2022driverShortage}. Labor issues significantly increased container congestion at the port, prompting Union Pacific to suspend service at all inland rail ramps serving the complex in June 2023 \citep{LaRocco2023UPpaused}. This suspension was intended to encourage shippers to divert their cargo to alternate ports and facilitate uninterrupted rail operations to and from the ports. Furthermore, trucks were denied access to the port terminal because of the slow productivity. Beyond the workforce shortage,  e-commerce trends, globalized supply chains, pandemic impacts, congested highways in California, limited network connections and space, environmental concerns, regulations, and limited public-private collaboration contribute to transportation bottlenecks \citep{UNCTAD2022Bottleneck}.

Given these factors, it is critical to understand and improve port operations, which are vital for international trade.
The global trade landscape is complex and involves numerous stakeholders, intricate processes, and interdependencies. Simulation, crucial across various sectors like healthcare, transportation, and energy, can accurately model these complex systems.

While extensive simulation literature exists on port operations and associated transportation systems, none has delved into the intricate interplay between operational dynamics and resource agents caused by fluctuations in workforce availability (see Section \ref{Literature review}). In this study, we focus on optimizing a modified Intermodal Freight System to enhance cost and throughput efficiencies. Our proposed system aims to streamline operations by eliminating the necessity for container sorting at the ports. Instead, we plan to utilize rail transport for the direct shipment of containers to distribution centers located across four states — California (CA), Utah (UT), Arizona (AZ), and Nevada (NV). Sorting operations are conducted at these distribution centers instead of the conventional method of sorting at the on-dock rail terminal, near-dock rail terminal, or near-port satellite rail terminals (see Section \ref{Problem Description} for details).

The challenges facing our proposed system include the limited hinterland connections of rail due to the absence of intermodal capability at logistics centers. Additionally, this transition necessitates a shift from trucking to other shipping modes. The proposed system also redefines the role of distribution centers, transitioning from cargo transfer and value-added services to acting as intermediaries for rail access and railcar sorting. We believe that adopting our proposed handling strategy will decrease overloaded sorting activities and congestion at port areas, and reduce reliance on truck transportation. Our primary objective is to conduct a comprehensive evaluation of the efficiency and effects of this proposed system, taking into account economic and societal factors, particularly in light of demand and workforce fluctuations.

We present the formal problem definition in Section \ref{Problem Description}. In Section \ref{Literature review}, we present a comprehensive literature review. Section \ref{Model Description} describes the model used. In Section \ref{Application and Results}, we cover the application and results, including data collection, parameter justification, simulation modeling, and findings. Finally, Section \ref{Conclusion} is dedicated to summarizing our work and discussing potential avenues for future research.

\section{Problem Description}
\label{Problem Description}

\begin{figure}[htpb] 
    \centering
    \includegraphics[width=0.7\textwidth]{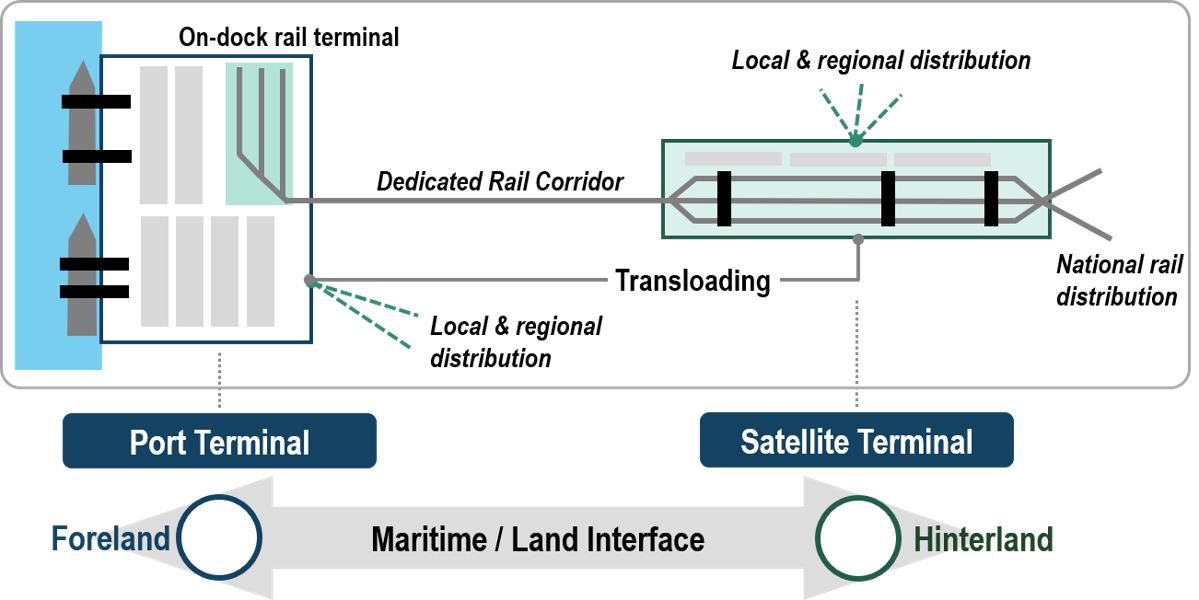}
    \caption[The insertion of a satellite terminal in port operations, adopted from]{The insertion of a satellite terminal in port operations, adopted from \citet{Rodrigue2020IntermodalTerminals}.}
    \label{fig:OnDock_NearDock_Terminals}
\end{figure}

In this section, we provide a brief overview of the operations and current practices for handling inbound containers. After being unloaded from vessels at the ports, containers are transferred to either the container terminal yard, rail terminal, classification yard, near-dock facility, or satellite terminal before being further transported via alternative modes to inland distribution areas, as illustrated in Figure \ref{fig:OnDock_NearDock_Terminals}. This process involves determining container movement and placement within those locations, temporary storage until sufficient quantities for a unit train are met, container sorting, transferring cargo into domestic containers, consolidation, deconsolidation, transloading, or occasionally providing value-added services.

The container classification, also known as the sorting process, involves categorizing railcars based on various criteria such as railroad company, loaded or unloaded status, destination, car type, and the need for repairs. Subsequently, railcars are grouped to compose trains. Ideally, organizing railcars according to their destination enables trains to optimize efficiency by avoiding unnecessary stops at multiple intermediary rail yards. However, this efficiency is often hindered by limited space and overactivities at and around the port area, resulting in congestion and bottlenecks that increase costs.

Considering that conventional approaches to expanding capacity, such as enlarging port facilities or upgrading equipment for efficiency, may face limitations due to development constraints, the utilization of nearby facilities emerges as the favored solution. Specifically in the U.S., container unit trains frequently exceed the length of container port terminals. Consequently, cargo segments are consolidated at on-dock rail facilities and subsequently transported to satellite rail terminals nearby in the hinterland for complete unit train assembly \citep{Rodrigue2022containerTerminal}. Satellite rail terminals are connected to port terminals via either rail shuttle services or, more commonly, truck drayage services \citep{Rodrigue2020IntermodalTerminals} as depicted in Figure \ref{fig:OnDock_NearDock_Terminals}. 

The current practice of employing satellite rail terminals as complementary to on-dock facilities at ports for classification aims to alleviate congestion, improve throughput, and facilitate the expansion of operational zones by diverting some freight traffic, cargo storage, and sorting activities to less crowded regions. Two primary shipment sorting methods are utilized at on-dock facilities: block swap and no-sort shuttle trains. In the block swap process, container car blocks are initially sorted at the on-dock facility for inland destinations, forming full-length trains. These trains are then transported to inland facilities to assemble destination-specific unit trains. Conversely, in the no-sort shuttle trains method, unsorted full-length trains are formed at the on-dock rail terminal and then transported to inland facilities for sorting into destination-specific unit trains. Both methods necessitate ample yard space for sorting operations.

To complement rail transport for non-local distribution, trucks are also utilized to retrieve containers from port terminals and transport them to distribution centers for cargo transfer into domestic containers. Subsequently, the trucks proceed to satellite rail terminals, where the cargo is transferred onto freight trains \citep{Rodrigue2022containerTerminal}. Regarding containers designated for local and regional distribution, typically assigned to trucks rather than railroads, they often linger at the port for prolonged durations, awaiting unloading and reloading into domestic containers. This process can take place at on-dock terminals or distribution centers. Once completed, trucks transport the containers to their destinations.

Despite the implementation of satellite terminals and the alteration in the sorting process, congestion persists primarily because all operations are still concentrated near the port area. In addition, this extensive process consumes significant time and resources and exacerbates shortages of drivers and port operators.

\section{Literature review}
\label{Literature review}

We explore the utilization of simulation methodologies to enhance supply chain management, specifically targeting the optimization of freight transportation processes at container terminals. The versatility of this method extends to tackling a wide array of challenges, including scheduling, layout design, work configuration, and the integration of cutting-edge technologies such as robotic vehicles, communication systems \citep{cimino2017evaluating}, and digital twins \citep{neugebauer2024digital}. Key outcomes include operation costs, queue length, waiting times, service time, time in systems, throughput levels, and resource utilization.

To start with, we refer the reader to \citep{dragovic2017simulation}  for a comprehensive exploration of simulation in port operations. The authors highlighted the predominant focus on container terminal applications within the realm of simulation in port operations. Out of 219 papers analyzed, 75.8\% of the research was dedicated to this specific area, indicating its continued significance in port simulation studies. 
Examples of subareas within container terminal applications include intermodal operations, automation, storage policies, logistics planning, integration of simulation and optimization models, and operational policies.

\citet{tao2015simulation} developed a simulation optimization method for dispatching port internal vehicles among multiple container terminals during loading or unloading. Their approach aims to reduce ship serving time and minimize vehicle empty traveling,  enhancing overall operational efficiency within the port. \citet{gharehgozli2017simulation} investigated the twin operations of automated stacking cranes at a container terminal, examining various strategies for locating storage areas, determining optimal container quantities, and devising efficient working schedules.

In another work, \citet{altayeb2017multi} employed multi-level modeling and simulation to tackle the complexities of port management. The first model level concentrates on essential port operations, encompassing distribution, port utilization, total time ships spend, and queuing dynamics. In parallel, the second model level delves into simulating the port's specific processes for container and U.S. Liquefied Natural Gas (LNG) loading and unloading operations. \citet{zhou2020container} optimized container handling operations by using simulation to evaluate the impacts of container reshuffling and integrating it with a Mixed Integer Linear Programming (MILP) model to address the storage space allocation problem. The majority of papers in this field focus on improving port and terminal processes within the port and marine terminals.  

Papers going beyond a singular focus on port operations often consider broader aspects such as environmental impact, sustainability, and supply chain resilience. For instance, \citet{castilla2020simulation} focused on transshipment operations, formulated as a quay crane scheduling problem, which is complemented by simulation to replicate the schedules of quay cranes. The authors proposed an algorithm to solve this optimization-simulation model, incorporating constraints related to the availability of internal delivery vehicles, and addressing uncertainties inherent to port operations, such as perturbations and interferences among quay cranes. \citet{muravev2021multi} developed an agent-based discrete-event simulation model to assess mitigation options for operational uncertainties and disturbances at intermodal terminals, such as adverse weather conditions, frequent changes in vessel schedules, and equipment breakdowns.

In addition, \citet{izaguirre2021climate} investigated the mitigation of risks posed by climate change to global maritime trade within the world port network. The research conducted by \citet{do2016simulation} aimed to reduce emissions from idle waiting trucks during the import container pick-up process by enhancing crane operations and optimizing truck time slot assignments. While there are many papers discussing uncertainty in port operations, none have thoroughly explored the complex interaction between operational dynamics and resource agents resulting from fluctuations in workforce shortages. %, including external vehicles and drivers.

In addition to utilizing simulation to model port processes and operations, some research focuses on understanding the broader cooperation within the supply chain system. Examples include, \citet{ramirez2017impact} highlighted the significance of coordination between seaside and hinterland terminals, with a specific focus on enhancing truck appointment system schedule policies as a means to improve this coordination. 
\citet{fedtke2017comparison} proposed four different handling systems designed for rail-rail transshipment yards, which serve as central hub nodes within a railway network for container consolidation between different trains. Each handling system entails decision-making regarding necessary equipment, fleets, and operational procedures aimed at enhancing the efficiency of the sorting system and scheduling.

The paper by \citet{morra2019case} employed AnyLogic software to simulate bulk material flow in an intermodal transportation system from mine to maritime terminal via trucks, trains, and automated conveying. The primary focus of the model was to assess the feasibility of the system and ensure synchronization among subsystems in terms of capacity and timetable. 
\citet{azab2020simulation} investigated the management of external truck appointments and explored fostering collaboration between trucking companies and gate operations to optimize truck turnaround time, workload distribution, and yard operations. 

In subsequent pioneering research, \citet{jacyna2020road} addressed the Road Vehicle Sequencing Problem for import container loading, particularly examining how random truck availability affects handling equipment operation time and the number of equipment required at a railroad inland intermodal terminal, aiming to minimize handling equipment operation time. This truck availability issue represents a typical disruption scenario where vehicles scheduled for loading may experience delays, resulting in interruptions to the loading schedule. As evidenced by several research efforts focusing on optimizing operations within inland terminals and investigating the impact of external trucks, there remains a gap in exploring strategies for identifying new locations for inland terminals that can accommodate additional freight containers from ports, integrating intermodal features encompassing both rail and truck transport.

This work has three contributions. Firstly, while previous studies have primarily focused on the operational performance of container terminals, our research extends beyond that by examining alternative locations for container terminals. This approach aims to reduce congestion costs and enhance the throughput of containers from ports. Secondly, concerning system uncertainty and disturbances, most studies have explored the impact of natural disasters or system failures. In contrast, our work investigates the significance of workforce shortages, which often hinder vessel entry operations, prolong storage times at ports, and result in significant supply chain delays. Finally, the research stream has largely overlooked opportunities for optimizing rail system utilization. We aim to address this by encouraging the shift in transportation modes. Through our proposed handling strategies, we seek to evaluate the performance of these new systems, achieve resource usage equilibrium, and enhance resource management under various conditions.

\section{Model Description}
\label{Model Description}

We aim to enhance containerized freight throughput and resource utilization while minimizing transportation costs and port congestion. We select the SPPC as a real-world case study. Using simulation, we create a container handling strategy that decreases reliance on the port's classification yard and reduces train assembly time, especially in response to varying demands and workforce fluctuations. The inbound transportation network starts when containers arrive, after which they are transported either by rail or truck to logistics centers  in CA and neighboring states: NV, UT, and AZ. We make the following set of assumptions:
\begin{itemize}
    \item Containers are directly transferred from cargo ships to trains and transported to logistics centers. This deviates from current port operations, where containers are stored in the classification yard for an extended period until they can be grouped with others destined for the same location. The classification process in our proposed handling strategy occurs at intermodal logistics centers. This approach relocates processes that consume significant time and space to other areas with less congestion, thereby alleviating congestion around the port area.
    \item Due to our new system, which does not require classification before loading onto trains, traditional high fixed costs associated with railroad loading are assumed to be reduced or eliminated. Our loading process does not require extra steps, such as unloading from ships and keeping goods in container yards for sorting, loading sorted containers onto railcars, or sorting railcars to form a train. Therefore, the process is expected to be faster and more cost-effective. 
    \item In our model, trains operate exclusively with full loads, while trucks handle both full truckload (FTL) and less-than-truckload (LTL) shipments. FTL refers to shipments that occupy an entire truck, while LTL involves combining multiple smaller shipments from different customers into a single truck. Trucks are solely responsible for transporting freight from warehouses to destinations.
    \item The demand for U.S. states is established using the 2021 weight of commodity data for waterway imports to California and distribution to all states, sourced from the Freight Analysis Framework Version 5 (FAF5) dataset \citep{BTS2022}. All containers, whether directly from the port or via warehouses, must meet destination demand.
\end{itemize}

\begin{figure}[htpb]   
{
        \centering\includegraphics[width=0.8\textwidth]{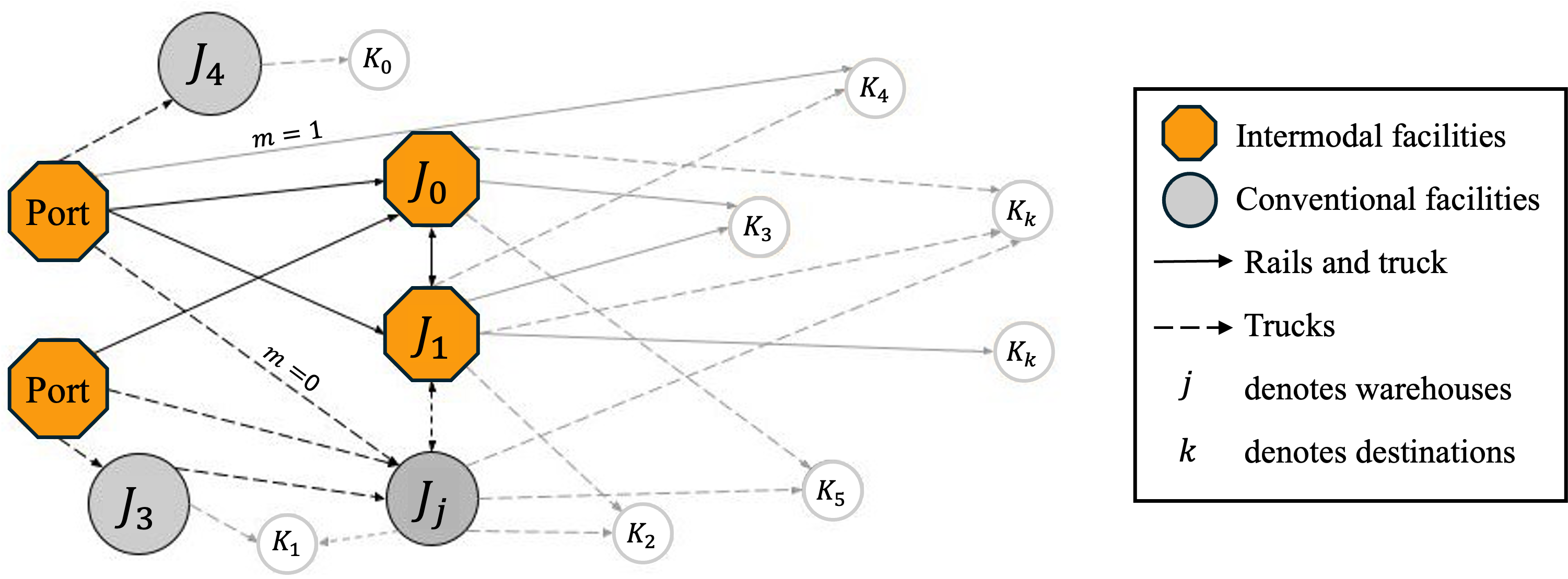}
		\caption{Proposed network of the southwest supply chain.}
        \label{fig:proposedNetwork}
}
\end{figure}

Figure \ref{fig:proposedNetwork} represents our proposed network from the port complex to warehouses ($J_j$) and destinations ($K_k$). Solid lines indicate connectivity by both truck and rail systems ($m=1$), while dashed lines denote connectivity by only truck systems ($m=0$). Octagonal nodes represent ports and logistics centers with intermodal facilities, while shaded circular nodes represent logistics centers without intermodal facilities. The uncolored nodes illustrate destinations, which may or may not be equipped with intermodal facilities. These destinations can be other regional logistics centers, local warehouses, companies, and retailers. The dimly drawn lines and nodes constitute complete networks that extend beyond our decision problems.

\section{Application and Results}
\label{Application and Results}

In this section, we will delve into the methodology of data collection and justify the chosen parameters, followed by modeling, results, and experiments. The model and data are accessible to the public in an online repository: \url{https://github.com/SaritaRkp/Simulation-model_SPPC}.

% \begin{figure}[htpb]   
%         \centering
%         \includegraphics[width=0.8\textwidth]{figs/infographic.jpg}
% 		\caption{Project Description and Tasks}
%         \label{fig:infographic}
% 	\end{figure}

\subsection{Data Collection and Parameters Justification}

The decisions regarding the intermodal logistics centers' location and shipping assignments depend on several factors, including shipping costs, distance, logistics center and vehicle capacity, and the potential congestion that transportation may cause. In our objective function, we include the following operational costs per mile: \$1.96 for trucks with congestion, \$1.65 for trucks without congestion, \$110.79 for long-haul rail, based on data generated from the research conducted by  \citet{RattaLBCA}. These figures are originally represented in 2020 dollars per ton-mile, equating to \$251.24 for trucks and \$31.07 for rails. Thus, trucks are approximately 8 times more expensive than rail for long distances. According to \citet{Mahmudi2006Railtruckcost}, the shipping costs for short-distance rail transport, where rail becomes less cost-effective compared to trucks, are roughly 2-3 times higher than using trucks.  Therefore, assuming these findings, the operational cost for short-haul rail could be around 20 times more expensive compared to long-haul rail, equivalent to \$2,380.55 per mile.
 We also assume a cost of \$900,000 per facility for upgrading a warehouse to an intermodal facility. For parameters related to tonnage and vehicle capacity, we assume a truck capacity of 8.475 tons, considering load factor and empty movements per vehicle, and a train capacity of 4,079 tons per train for Class I railroads \citep{RattaLBCA}. 

The challenge in the data collection process lies in gathering warehouse capacity data. We collect real-world GIS data, including capacity and functionality, from our targeted states: CA, NV, AZ, and UT. Since this data is not publicly available, we justified our estimates using several factors such as warehouse space, number of rail doors, storage capacity, and annual lift. In cases where specific details were unavailable, we assumed the warehouse was relatively small, handling only 0.07\% of all inbound freight from CA. This assumption is designed to ensure that the cumulative capacity of all our warehouses is slightly less than the total inbound freight, accounting for potential oversights in modeling certain warehouses.

Originally, we started with 570 locations. However, after encountering issues with the simulation exceeding the agent limit in the AnyLogic PLE version, we opted to remove smaller locations, resulting in 470 locations. Despite this reduction, we continued to face computational limitations. To address this, we further grouped adjacent warehouses with rail capability and those without rail capability, combining the capacity of each cluster. This led to a total of 245 locations. We proceeded with experiments on this dataset until we encountered challenges in optimizing how trains select the next warehouse for cargo drop-offs.

Initially, random selection was sufficient. However, when we prioritized the nearest available warehouse for train cargo drop-offs, we faced computational limits. To address this, we combined more warehouses, reducing to 35 locations while maintaining the original network capacity.
 
% \begin{figure}[htpb]   
%         \centering		\includegraphics[width=0.9\textwidth]{figs/WarehouseDatabase.png}
% 		\caption{Warehouse Database}
%         \label{fig:warehouse database}
% 	\end{figure}

For customer data integration, the process is straightforward. We utilized the FAF5 databases established by BTS \citep{BTS2022}, filtering only the imported data that reached CA via the waterway. From this, we determined the tonnage received by the remaining 50 states and Washington DC for those shipments and calculated the percentage accordingly. Upon reviewing the percentages, it became evident that numerous states received only a minimal amount of these shipments. Consequently, we eliminated those states, resulting in a refined list of 18 states that remain the primary customers for the targeted inbound freight. These states collectively account for 93.23\% of the total inbound freights.

% \begin{figure}[htpb]   
%         \centering		
%         \includegraphics[width=0.9\textwidth]{figs/CustomerDatabase.png}
% 		\caption{Customer Database}
%         \label{fig:customer database}
% 	\end{figure}

\begin{figure}[htpb]   
{
        \centering		\includegraphics[width=.75\textwidth]{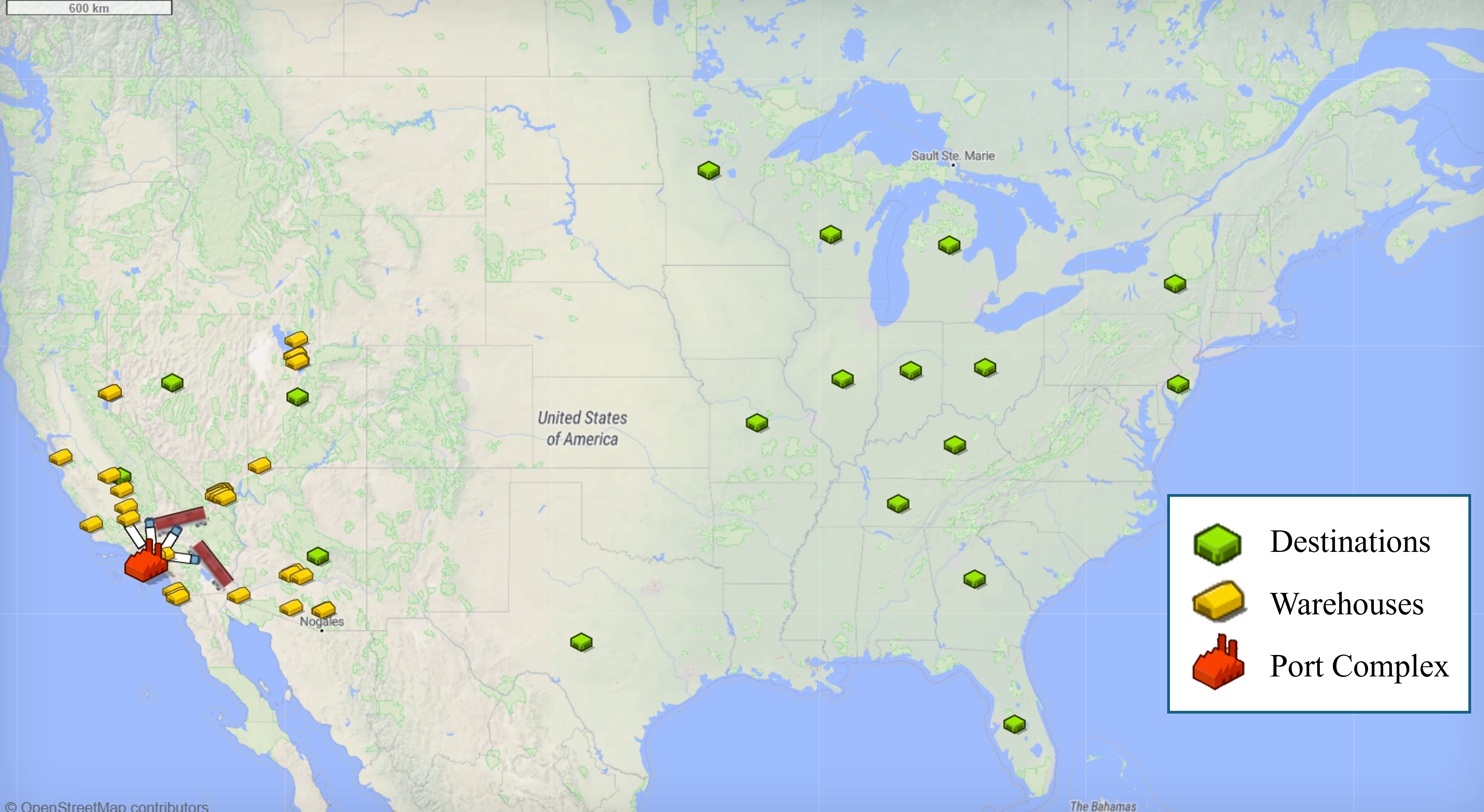}
		\caption{Network topology.}
        \label{fig:networkTopology}
}
\end{figure}

For the port database, our focus centers on the port complex, given its status as one of the busiest ports in both CA and the entire U.S. Figure \ref{fig:networkTopology} illustrates our network topology, including destinations, warehouses, and the port. We have two modes of transportation: trains and trucks. For trucks, there are two types of agents. The first type consists of trucks originating at the port and moving towards warehouses. The second type consists of trucks originating at warehouses and traveling between warehouses and destinations.

% \begin{figure}[htpb]   
%         \centering		\includegraphics[width=0.6\textwidth]{figs/PortDatabase.png}
% 		\caption{Port Database}
%         \label{fig:port database}
% 	\end{figure}

\subsection{Modelling}
We utilized AnyLogic software version 8.8.5 for our agent-based simulation model, incorporating nine types of agents: Main, Customer, Order, PortLA, PortTruck, Shipment, Train, Warehouse, and WarehouseTruck. We used Java for AnyLogic and Python for ANOVA tests. Before the simulation, we determine the number of new warehouses to be equipped with intermodal facilities (P) for warehouses. This decision, made at the start of the year,  persists for consecutive periods (i.e., the entire year). Subsequently, the simulation proceeds to determine transportation operations, which include vehicle routing, tonnage allocation for each warehouse, and vehicle assignment.  We set the number of agents at the port complex to be fifteen trucks and two trains, and the sum of the number of trucks at all warehouses to be 280 trucks. 
 
% \begin{figure}[htpb]   
%         \centering		\includegraphics[width=1\textwidth]{figs/main3.png}
% 		\caption{Main Agent's Parameters and Variables}
%         \label{fig:main agent}
% 	\end{figure}

For the port agent, the shipment agent arrives at the source with a total tonnage divided by twelve months, and it follows an interarrival time of one month. All shipments wait in the queue, checking if there is an available truck or train to load the shipment onto. Depending on the shipment size, if it is substantial enough to form a full train, we prioritize sending it via train. In this case, if the shipment exceeds the train's capacity, the excess is split and placed back into the queue. The train then proceeds to the warehouse. 
As mentioned earlier, for trucks, both LTL and FTL options are allowed.  

% \begin{figure}[htpb]   
%         \centering		\includegraphics[width=1\textwidth]{figs/PortLA.png}
% 		\caption{Port of Los Angeles Agent}
%         \label{fig:portLA agent}
% 	\end{figure}

% Figure \ref{fig:Train agent} illustrates the train's behavior using a statechart to model its operations.

To model the trains' operations, we employ a statechart. When the port receives a message requesting cargo dispatch based on the monthly demand, the train initiates a decision-making process to identify warehouses with available capacity, then the train agent randomly selects one warehouse and delivers the cargo. Following this, another decision ensues: if more cargo remains on the train, it repeats the decision process to deliver cargo to the next warehouse. At this point, the train can choose between delivering the cargo to a randomly selected warehouse or to the nearest available warehouse in proximity to the one where the cargo was recently dropped. Once the train is emptied, it returns to the port.

% \begin{figure}[htpb]   
%         \centering		\includegraphics[width=0.8\textwidth]{figs/Train.png}
% 		\caption{Train Agent}
%         \label{fig:Train agent}
% 	\end{figure}

The port truck agents' behavior closely resembles that of the train agents, with two key distinctions: there is no consideration for selecting the nearest warehouse, and an additional factor is introduced—the uncertainty of driver shortage. Trucks deliver their entire load to a single warehouse, given their relatively small average load of eight tons. Regarding driver shortage, we introduced an extra stage called 'ReplenishDriver' triggered for long hauls  (over 250 miles, such as traveling from the port to outside CA), causing driver dissatisfaction that can impact driver retention and shortage. We assume a uniform probability distribution for driver shortage in three cases: 0-24 hours, 0-72 hours, and 0-200 hours.
The warehouse truck operates between warehouses and destinations, sharing the same characteristics as the port truck agent, except for the ReplenishDriver process, as it falls outside the scope of our proposed handling system focus.

At each warehouse, a warehouse agent is responsible for defining how to load the inventory into their fleet of trucks, given there is an available truck. If the inventory exceeds the truck capacity, the agent loads the truck until it is full; if the inventory is less than the truck capacity, the agent loads the truck until the inventory is empty. Afterward, the agent returns to the queue to process the next order, while the loaded truck proceeds to the shipping and delivery process destined for a specific location.

Finally, we perform verification and validation of our model to ensure it accurately represents the problem and assumptions. For verification, we initially focused on a shorter one-month period to rigorously test the model's functionality. We scaled down the annual tonnage to monthly demand and simulated monthly shipments starting each month, mirroring the setup repeated twelve times annually in our year-long model.

Verification involved various metrics and experiments. We measured the flow of goods in and out of each node and simulation block to ensure precise aggregation, division, and assignment of flows. We assessed the impact of varying fleet capacities on system performance and adjusted agent behaviors, like changing warehouse selection criteria from random to proximity-based, to observe outcome changes. Further, model accuracy was verified by comparing manually computed distances and costs, among other procedures.

For validation, we focused on key operational metrics. The average dwell times of containers waiting for pickup at SPPC were measured at 5.8 days for rail pickups and 2.5 days for truck pickups. These findings closely aligned with real-world data showing average dwell times of 4.3 days for rail and 3.2 days for truck pickups \citep{Biggar2023DwellTime}, validating the model's ability to simulate operational realities effectively.

\subsection{Results and Discussion}
We conducted three experiments on the main agent, encompassing simulation, parameter variation, and optimization, as detailed in the following subsections.

\subsubsection{Installing Intermodal Equipment Simulation}

\begin{figure}[htpb]   
        \centering		\includegraphics[width=.75\textwidth]{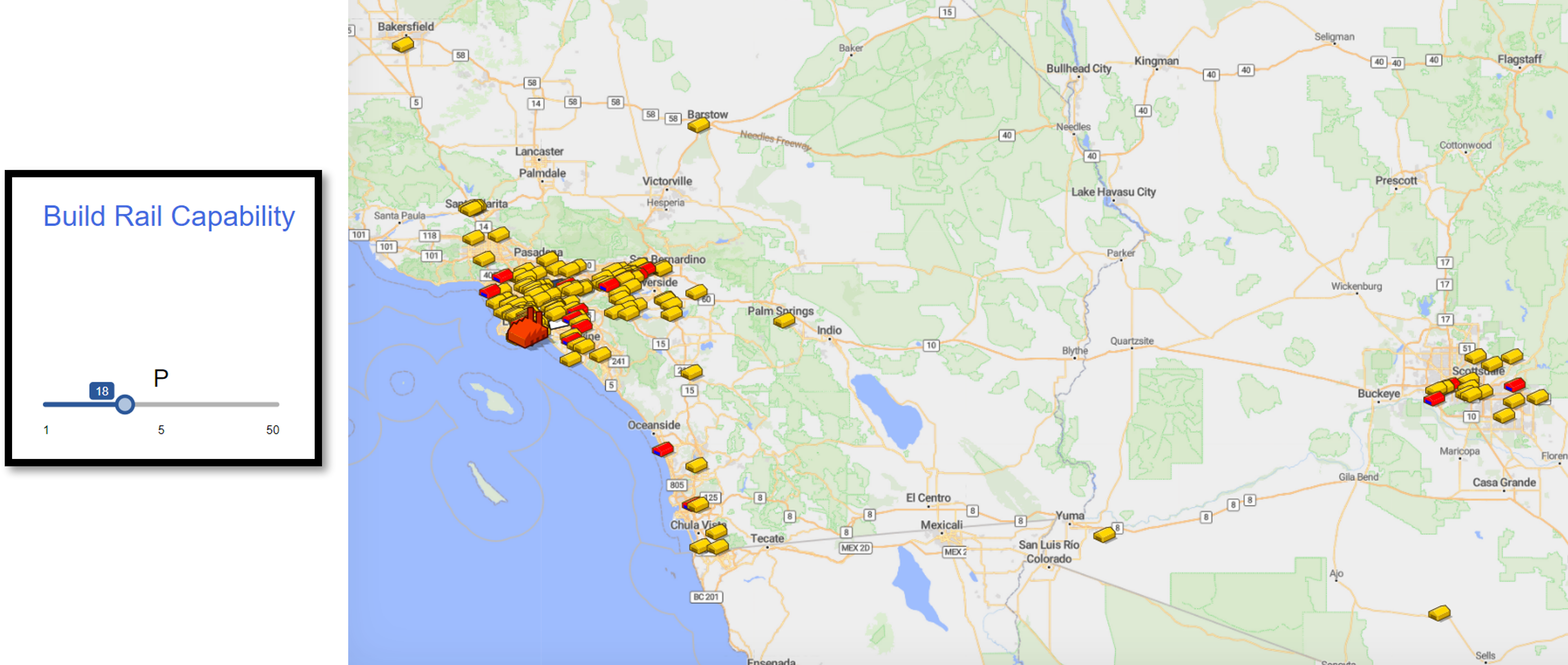}
		\caption{A demonstration of installed intermodal equipment.}
        \label{fig:addP Simulation}
	\end{figure}

Before initiating the simulation, we can define the number of warehouses to equip with intermodal capability (P). A slider is used to adjust the number of P; the selected ones are displayed in red. Original warehouses, regardless of their intermodal capability, are depicted in yellow. In the model, as shown in Figure \ref{fig:addP Simulation}, it is observable that certain warehouses turn red, indicating that these warehouses originally lacked intermodal capability but were randomly chosen for installation with intermodal facilities before the simulation began. The output of the simulation includes costs, the amount of unmet demand, and resource utilization.

\subsubsection{Parameter Variation Experiment}

We conducted five different model runs using Parameters Variation Experiments with random seeds. For the first setting,  we completed only 69 runs due to reaching the maximum number of agents, which terminated the execution. However, the remaining model settings successfully achieved our goal of 100 iterations.

The evaluation of the model's performance is based on three factors: unmet demand, resource utilization, and total costs (see Table \ref{tab:computational result}). Each cell contains two numbers: the first number represents the mean value, while the number inside the parenthesis indicates the minimum value observed across those iterations.

Surprisingly, in terms of unmet demand, the model performed best when the train dropped cargo at random warehouses (refers to Random WHS in Table \ref{tab:computational result}). This model resulted in the overall warehouse network's freight distribution speed, leading to the discovery of the minimum unmet demand among all 569 iterations. This outcome is likely attributed to fleet utilization at each warehouse. When multiple trains drop cargo at a cluster of warehouses using the nearest warehouse selection (referred to as N-WHS), it can lead to high fleet utilization at certain times. This situation persists until no available trucks can serve those warehouses, consequently increasing waiting times. As a consequence, warehouses far away from that cluster become nearly empty during that period, with the fleet having approximately 80\% availability.

It is important to note that fleet utilization at warehouses is not included in the table, as adding this data would lead to 175 numbers (35 warehouses * 5 model settings).  Hence, we will provide a concise explanation: of the 35 warehouses, roughly half utilized between 80-98\% of the resource pool by the year's end, while the other half utilized only 10-20\%, mainly due to the selection of the nearest warehouse. The Resource column in the table only represents the fleet (truck) and rail at the port, not the fleets at individual warehouses. Nevertheless, the nearest selection of the warehouse for trains to drop cargo proves beneficial for resource utilization at the port, as both the truck and rail fleets performed most efficiently and became available more quickly than the other five settings. 

In the four last settings with driver replenishment distribution (DRD) at different cases following a uniform distribution of 0-24 hours, 0-72 hours, and 0-200 hours, we observe that as the time to replenish a driver increases, the fleet becomes busier, and unmet demand also increases. In the last setting (DRD and N-WHS), which combines a DRD of 0-200 hours with the nearest warehouse selection, the results are similar to before. The nearest warehouse selection helps improve resource utilization at the port area, yet it exacerbates the unmet demand problem for the same reasons explained earlier.

For the cost factor, the results do not explicitly represent the performance of the model, as they do not exhibit a clear trend. This lack of clarity may be attributed to the model's complexity, involving numerous agents with substantial uncertainty from each. It is possible that more than 100 interactions at each model setting would be necessary for a more comprehensive understanding, but this is constrained by the maximum number of agents.

After obtaining the results, we conducted one-way ANOVA tests to analyze the differences among all six settings. Specifically, we tested four factors: fleet utilization, train utilization, cost, and unmet demand. All ANOVA tests yielded p-values very close to 0, which is less than 0.05, indicating a rejection of the null hypothesis. This suggests significant differences in performance among our six model settings.
                      
\begin{table}[htb]   
\centering
\caption{\label{tab:computational result}Computational result}
\begin{tabular}{p{4cm}|p{1.6cm}|p{2.7 cm}|p{2cm}|p{2.5cm}} 
Models & Iterations & Unmet Demand & Resource & Costs \\\hline
Random WHS & 69 & 424,263.8 (413,836) & fleet: 34.8\% rail: 35.4\%   & 2,695,916.8 (2,605,807.7)\\
N-WHS & 100 & 516,965.4 (471,941.1) & fleet: 22.8\% rail: 22.7\%   & 2,178,682.3 (2,053,240) \\
DRD (0, 24) & 100 & 486,851.3 (457,927.4) & fleet: 37.5\% rail: 37.3\% & 2,683,171.6 (2,445,183.1)\\
DRD (0, 72) & 100 & 517,976 (481,076) & fleet: 40.8\% rail: 38.5\% & 2,667,781.2 (2,361,362.4)\\
DRD (0, 200) & 100 & 520,124.4 (496,429.3) &  fleet: 46.5\% rail: 38.6\%  & 2,665,136.9 (2,493,952.4)\\
DRD (0, 200) \& N-WHS  & 100 &  551,250.8 (542,458.5) & fleet: 33.2\% rail: 24.2\% & 2,103,508.2 (1,884,490.2) \\\hline
\end{tabular}
\end{table}

\subsubsection{Optimization of Intermodal Equipment Installation (with random seeds)}
In the optimization experiment on the model with the selection of the nearest warehouse and Driver Replenishment Distribution set to a uniform discrete (0, 72), our objective is to minimize the total cost. We varied the parameter P, representing the number of warehouses equipped with intermodal facilities, ranging from 0 to 10, with a step of 1. Despite the limitation on the maximum number of agents, we conducted several experiments and found that installing four intermodal facilities at existing warehouses led to an enhancement in total costs compared to the 100 iterations of the parameter variation experiment conducted under the same settings. The new cost is \$2,100,413, as opposed to the minimum cost observed in the 100 iterations, which was \$2,361,362. However, it is important to note that this conclusion might not always hold, given the model's inherent uncertainty stemming from the numerous agents involved as explained in the previous subsection.

\section{Conclusion}
\label{Conclusion}

Our proposed system necessitates substantial changes in shipping modes and railroad operations. It aims to enhance logistical efficiency by relocating handling strategies and classification processes away from port areas, mitigating workforce shortages by balancing resource utilization and engaging independent truckers outside CA, while also leveraging the cost-efficient rail system. We conducted three experiments on the main agent: simulation, parameter variation, and optimization. The model's performance evaluation revolves around three metrics: unmet demand, resource utilization, and total costs. Our proposed system enhances resource utilization at the port complex, particularly when unsorted full-length trains are expedited to the nearest available warehouses regardless of destination. However, this approach can lead to an imbalance in resource usage across warehouses. By comparing N-WHS to a randomized selection method, we observed that randomly selecting warehouses for cargo offloading from rails can improve the overall distribution speed across the warehouse network, ultimately minimizing unmet demand across all 569 iterations. Regarding cost analysis, the results lack a clear trend, likely due to the model's complexity and the inherent uncertainty associated with multiple agents. Nonetheless, in comparing the scenario of the nearest warehouse selection with DRD set to a uniform discrete (0, 72), the optimization experiment highlights that installing intermodal capability at four warehouses yields the most favorable cost outcome among all 100 iterations.

Enhancing model accuracy is crucial for its reliability and effectiveness. Firstly, our current focus is on the POLA and the POLB as the main hubs for CA inbound freight handling, given their prominence among the 12 ports in CA. Expanding our model to incorporate the dynamics of multiple ports can enhance its representativeness and applicability. Secondly, the demand utilized in our model is currently derived from historical data of the year 2021. To better reflect real-world scenarios, we should incorporate demand uncertainty in the future, providing a more realistic evaluation of our model's performance under varying conditions. Additionally, our results and conclusions have highlighted imperfections in the train agent, particularly concerning the randomly selected and nearest selection processes. Developing a more robust and optimized process for train assignment is essential for improving the overall efficiency of our model. Finally, regarding implementation, our current model lacks emphasis on warehouse operations (e.g., storage, packing, labeling), specific commodity-based warehouse classification, and cooperation with associated sectors. Moreover, we have not fully addressed equipment usage impacts, such as longer maritime container travel to inland warehouses and congestion from clearing gate procedures. These challenges require further study for a comprehensive implementation of freight handling and classification processes.

\end{document}